\documentclass[12pt,letterpaper]{article}
\pdfoutput=1
\usepackage{jheppub}
\usepackage{amsfonts, amsthm}
\usepackage[english]{babel}
\usepackage[utf8]{inputenc}
\usepackage{slashed}
\usepackage{mathrsfs}
\hypersetup{unicode}

\newcommand{\eq}{\begin{equation}}
\newcommand{\feq}{\end{equation}}
\newcommand{\eqn}{\begin{eqnarray}}
\newcommand{\feqn}{\end{eqnarray}}

\title{BPS black holes in a non-homogeneous deformation of the stu model of
$N=2$, $D=4$ gauged supergravity}

\author[a]{Dietmar Klemm,}
\author[b,c]{Alessio Marrani}
\author[a]{Nicol\`o Petri}
\author[a]{and Camilla Santoli}

\affiliation[a]{Dipartimento di Fisica, Universit\`a di Milano, and \\
INFN, Sezione di Milano, \\
Via Celoria 16, I-20133 Milano, Italy.}
\affiliation[b]{Centro Studi e Ricerche `Enrico Fermi', \\
Via Panisperna 89A, I-00184 Roma, Italy.}
\affiliation[c]{Dipartimento di Fisica e Astronomia `Galileo Galilei', \\
Universit\`a di Padova, and \\
INFN, Sezione di Padova, \\
Via Marzolo 8, I-35131 Padova, Italy.}

\emailAdd{dietmar.klemm@mi.infn.it}
\emailAdd{Alessio.Marrani@pd.infn.it}
\emailAdd{nicolo.petri@mi.infn.it}
\emailAdd{camilla.santoli@mi.infn.it}
\preprint{IFUM-1042-FT\\ \hspace*{\fill} DFPD/2015/TH/17}

\abstract{
We consider a deformation of the well-known stu model of $N=2$, $D=4$ supergravity, characterized
by a non-homogeneous special K\"{a}hler manifold, and by the smallest electric-magnetic duality Lie
algebra consistent with its upliftability to five dimensions.
We explicitly solve the BPS attractor equations and construct static supersymmetric black holes
with radial symmetry, in the context of $\text{U}(1)$ dyonic Fayet-Iliopoulos
gauging, focussing on axion-free solutions. Due to non-homogeneity of the
scalar manifold, the model evades the analysis recently given in the literature.
The relevant physical properties of the resulting black hole solution are
discussed.
}

\keywords{Black Holes, Supergravity Models, Black Holes in String Theory.}

\begin{document}
\maketitle
\flushbottom

\section{Introduction}

Black holes in gauged supergravity theories provide an important testground to address
fundamental questions of gravity, both at the classical and quantum level. Among the most
prominent of these are perhaps the problems of black hole microstates, uniqueness theorems,
or the attractor mechanism. In gauged supergravity, the solutions often (but not necessarily)
have AdS asymptotics, and one can then try to
study at least some of these issues guided by the AdS/CFT correspondence. On the other hand, black
hole solutions to these theories are also relevant for a number of recent developments
in fluid mechanics, high energy- and especially in condensed matter physics, since they provide the
dual description of certain strongly coupled condensed matter systems at finite temperature,
cf.~\cite{Hartnoll:2009sz} for a review. In particular, models similar to the one that we shall consider below,
containing Einstein gravity coupled to $\text{U}(1)$ gauge
fields and neutral scalars, have been instrumental to study transitions from Fermi-liquid
to non-Fermi-liquid behaviour, cf.~\cite{Charmousis:2010zz,Iizuka:2011hg} and references therein.

For these reasons, the construction of analytic black holes in gauged supergravity as well as
the exploration of their physics has been an active field of research recently, especially in four-dimensional
models with $N=2$ supersymmetry and Fayet-Iliopoulos (FI) gaugings, cf.~\cite{Duff:1999gh,Sabra:1999ux,Chamseddine:2000bk,Chong:2004na,Cvetic:2005zi,Bellucci:2008cb,Cacciatori:2009iz,
Chow:2010fw,Dall'Agata:2010gj,Hristov:2010ri,Klemm:2011xw,Donos:2011pn,Colleoni:2012jq,
Klemm:2012yg,Toldo:2012ec,Klemm:2012vm,Gnecchi:2012kb,Lu:2013ura,Halmagyi:2013qoa,
Chow:2013gba,Gnecchi:2013mja,Gnecchi:2013mta,Halmagyi:2013uza,Katmadas:2014faa,Gnecchi:2014cqa,
Halmagyi:2014qza,Erbin:2015gha} for an (incomplete) list of references. Although we are still
far from understanding the underlying general structure\footnote{By this we mean a possible gauged
supergravity analogue of the well-known fact that asymptotically flat black holes are typically given (in
the extremal limit) in terms of harmonic functions on a flat base space.} of such solutions (if there is any),
many important partial results have been obtained. These studies have also revealed some surprises,
like for instance the existence of so-called superentropic black holes, which have noncompact event
horizon but nevertheless finite area. These were first discovered in \cite{Gnecchi:2013mja}, and their
physics was further discussed in \cite{Klemm:2014rda,Hennigar:2014cfa,Hennigar:2015cja}.

Up to now, the construction and discussion of black holes in $N=2$, $D=4$ Fayet-Iliopoulos
gauged supergravity theories has been mainly limited to models where the vector multiplet scalars
parametrize a {\it symmetric} special K\"ahler manifold\footnote{For some notable exceptions
cf.~e.g.~\cite{Faedo:2015jqa}.}. Here we shall go one step further w.r.t.~the results that appeared in the
literature so far, by considering a non-symmetric (and even non-homogeneous) deformation
of the stu model, defined by the prepotential \eqref{eq:prepotential-pre}.
We will deal with a particular FI gauging of this model, that leads to a scalar potential with two
critical points corresponding to AdS vacua. One of these extremizes also the superpotential and
is thus supersymmetric, while the other vacuum breaks supersymmetry.

The remainder of the paper is organized as follows:
in section~\ref{Setup} we introduce some basics of the theoretical framework of
our investigation, namely $N=2$, $D=4$ supergravity coupled to $n_{V}$
vector multiplets, and its dyonic $\text{U}(1)$ Fayet-Iliopoulos gauging.
Then, in section~\ref{nH-STU} we focus on a specific model, whose three complex
scalars parametrize a non-homogeneous special K\"{a}hler manifold. At the
level of the prepotential, this is a one-parameter extension of the well-known
stu model \cite{Duff:1995sm,Behrndt:1996hu,Bellucci:2008sv}, and thus we call it a
non-homogeneous deformation of the stu model (nh-stu). In particular, respectively in
subsections~\ref{U-Duality} and \ref{Axion-Free-Geom}, we compute the
symplectic embedding of the electric-magnetic duality algebra, and we
present some axion-free geometric data.
In section~\ref{FI-Gauging} we perform a near-horizon analysis of the FI-gauged
system, in particular axion-free charge configurations, and for specific
choice of the dyonic FI gauging parameters.
A new, explicit BPS black hole solution for the FI-gauged nh-stu model is
presented in section~\ref{Solution}, and its physical properties are then
discussed in section~\ref{Phys-Discussion}.
The concluding section~\ref{Conclusion} contains some outlook and
considerations for future developments.

\section{\label{Setup}The setup}

We consider $N=2$, $D=4$ gauged supergravity coupled to $n_{V}$ Abelian
vector multiplets (for notation and general treatment, cf.~e.g.~\cite{Andrianopoli:1996cm}).
Besides the Vierbein $e_{\mu }^{a}$, the
bosonic field content includes the vectors $A_{\mu }^{\Lambda }$ enumerated
by $\Lambda =0,\ldots ,n_{V}$ (with the naught index denoting the
graviphoton), and the complex scalars $z^{i}$ where $i=1,\ldots ,n_{V}$.
These scalars parametrize a special K\"{a}hler manifold, i.e., an
$n_{V}$-dimensional Hodge-K\"{a}hler manifold that is the base of a
symplectic bundle, with covariantly holomorphic sections
\begin{equation}
\mathcal{V}=\left(
\begin{array}{c}
L^{\Lambda } \\
M_{\Lambda }%
\end{array}%
\right) \,,\qquad D_{\bar{\imath}}\mathcal{V}=\partial _{\bar{\imath}}%
\mathcal{V}-\frac{1}{2}(\partial _{\bar{\imath}}\mathcal{K})\mathcal{V}=0\,,
\label{eq:section}
\end{equation}%
where $\mathcal{K}$ is the K\"{a}hler potential and $D$ denotes the K\"{a}hler-covariant derivative. $\mathcal{V}$ obeys the symplectic constraint\footnote{The brackets represent the symplectic inner
product $\langle A,B\rangle
=A^{T}\Omega B=A_{\Lambda }B^{\Lambda }-A^{\Lambda }B_{\Lambda }$. \label%
{ftn:product}} $i\langle \mathcal{V}\,,\bar{\mathcal{V}}\rangle =1$, and it
is related to the holomorphic symplectic vector $(X^{\Lambda },F_{\Lambda
})^{T}$ by
\begin{equation}
\mathcal{V}=e^{\mathcal{K}/2}\left(
\begin{array}{c}
X^{\Lambda } \\
F_{\Lambda }%
\end{array}%
\right) \,.
\end{equation}%
The matrix $\mathcal{N}_{\Lambda \Sigma }$ determining the coupling between
the scalars $z^{i}$ and the vectors $A_{\mu }^{\Lambda }$ is defined by the
relations\footnote{%
In what follows we use the notation $\mathcal{I}=\mathrm{Im}\,\mathcal{N}$
and $\mathcal{R}=\mathrm{Re}\,\mathcal{N}$.}
\begin{equation}
M_{\Lambda }=\mathcal{N}_{\Lambda \Sigma }L^{\Sigma }\,,\qquad D_{\bar{\imath%
}}\bar{M}_{\Lambda }=\mathcal{N}_{\Lambda \Sigma }D_{\bar{\imath}}\bar{L}%
^{\Sigma }\,.  \label{defN}
\end{equation}
If a prepotential $F(X)$ exists, it is a homogeneous function of degree two
which allows to determine the lower part of the symplectic sections (\ref%
{eq:section}) and the matrix $\mathcal{N}$ in terms of $F$ itself, according
to
\begin{equation}  \label{eq:N}
\mathcal{V} = \mathrm{e}^{\mathcal{K}/2}\left(%
\begin{array}{c}
X^{\Lambda} \\
\partial_\Lambda F%
\end{array}%
\right)\,, \qquad \mathcal{N}_{\Lambda\Sigma} = \bar{F}_{\Lambda\Sigma}+2i%
\frac{\mathrm{Im}F_{\Lambda\Lambda^{\prime }} X^{\Lambda^{\prime }}\mathrm{Im%
}F_{\Sigma\Sigma^{\prime }}X^{\Sigma^{\prime }}} {X^{\Omega}\mathrm{Im}%
F_{\Omega\Omega^{\prime }}X^{\Omega^{\prime }}}\,,
\end{equation}
where $F_{\Lambda\Sigma}=\partial_{\Lambda}\partial_{\Sigma}F$.\\
The bosonic Lagrangian reads
\begin{equation}
\mathscr{L}=\frac{R}{2}-g_{i\bar{\jmath}}\,\partial _{\mu }z^{i}\partial
^{\mu }\bar{z}^{\bar{\jmath}}+\frac{1}{4}\,\mathcal{I}_{\Lambda \Sigma
}\,F_{\mu \nu }^{\Lambda }\,F^{\Sigma \mu \nu }+\frac{1}{8\sqrt{-g}}%
\,\epsilon ^{\mu \nu \rho \sigma }\mathcal{R}_{\Lambda \Sigma }\,F_{\mu \nu
}^{\Lambda }\,F_{\rho \sigma }^{\Sigma }-V_{g}\,,  \label{eq:Lagrangian}
\end{equation}%
with the special K\"ahler metric $g_{i\bar{\jmath}}=\partial _{i}\partial _{\bar{\jmath}}\mathcal{K}$.
The scalar potential is
\begin{equation}
V_{g}=g^{i\bar{\jmath}}D_{i}\mathcal{L}{\bar{D}}_{\bar{\jmath}}\bar{\mathcal{%
L}}-3|\mathcal{L}|^{2}\,,  \label{eq:scalar_pot}
\end{equation}
where the superpotential $\mathcal{L}$ is determined by the dyonic
Fayet-Iliopoulos (FI) gauging,
\begin{equation}
\mathcal{L}=\langle \mathcal{G},\mathcal{V}\rangle =e^{\mathcal{K}%
/2}(X^{\Lambda }g_{\Lambda }-F_{\Lambda }g^{\Lambda })\,,  \label{eq:L}
\end{equation}%
with FI parameters $\mathcal{G}=(g^{\Lambda },g_{\Lambda })$.\\
Since we are interested in static black holes with radial symmetry, we
employ the Ansatz
\begin{equation}
\mathrm{d}s^{2}=-e^{2U (r)} \mathrm{d}t^{2}+e^{-2U(r)}(\mathrm{d}%
r^{2}+e^{2\psi (r)}\mathrm{d}\Omega _{\kappa }^{2})\,,  \label{eq:metric}
\end{equation}%
where $\mathrm{d}\Omega _{\kappa }^{2}=\mathrm{d}\theta ^{2}+f_{\kappa
}^{2}(\theta ) \, \mathrm{d}\phi ^{2}$ is the metric on the two-surfaces $%
\Sigma =\{\text{S}^{2},\mathbb{E}^{2},\text{H}^{2}\}$ of constant scalar
curvature $R=2\kappa $, with $\kappa =\{1,0,-1\}$ respectively. Here the
function $f_{\kappa }(\theta )$ is given by
\begin{equation}
f_{\kappa }(\theta )=\left\{
\begin{array}{c@{\quad}l}
\sin \theta \,, & \kappa =1\,, \\
\theta \,, & \kappa =0\,, \\
\sinh \theta \,, & \kappa =-1\,.%
\end{array}%
\right.
\end{equation}%
The scalars are assumed to depend only on the radial coordinate $r$, $%
z^{i}=z^{i}(r)$, while the gauge fields should have an appropriate profile
to satisfy
\begin{equation}
p^{\Lambda }=\frac{1}{\text{vol}(\Sigma )}\int_{\Sigma }F^{\Lambda
}\,,\qquad q_{\Lambda }=\frac{1}{\text{vol}(\Sigma )}\int_{\Sigma
}G_{\Lambda }\,,
\end{equation}%
with $p^{\Lambda }$ and $q_{\Lambda }$ being the magnetic and electric
charges associated to the black hole and $G_{\Lambda }$ denoting the dual
field strength,
\begin{equation}
G_{\Lambda }=\frac{\delta \mathscr{L}}{ \delta\star\! F^{\Lambda }}\,.
\end{equation}
The symplectic invariant central charge is given by
\begin{equation}
\mathcal{Z}=\langle \mathcal{Q},\mathcal{V}\rangle \,,  \label{eq:Z}
\end{equation}
where we introduced the vector of magnetic and electric charges, $\mathcal{Q}%
=(p^{\Lambda },q_{\Lambda })$.

Following the procedure outlined in \cite{Dall'Agata:2010gj}, the previous
Ans\"{a}tze are plugged into the Lagrangian (\ref{eq:Lagrangian}) and give
rise to an effective one-dimensional action involving the scalar fields and
the warp functions $U(r),\psi (r)$,
\begin{equation}
\begin{split}
S_{1d}=& \int \mathrm{d}r\left\{ e^{2\psi }\left[U^{\prime 2}-\psi
^{\prime 2}+g_{i\bar{\jmath}}z^{i\,^{\prime }}\bar{z}^{\bar{\jmath}%
\,^{\prime }}+e^{2U-4\psi }V_{\text{BH}}+e^{-2U}V_{g}\right] -1\right\} \\
& +\int \mathrm{d}r\frac{\mathrm{d}}{\mathrm{d}r}\left[ e^{2\psi }(2\psi
^{\prime }-U^{\prime })\right] \,.
\end{split}
\label{eq:S_eff}
\end{equation}%
Here $V_{\text{BH}}$ denotes the so-called black hole potential \cite{FGK},
defined by
\begin{equation}
V_{\text{BH}}=-\frac{1}{2}\mathcal{Q}^{T}\mathcal{M}\mathcal{Q}\,,
\end{equation}
where
\begin{equation}
\mathcal{M}=\left(
\begin{array}{cc}
\mathcal{I}+\mathcal{R}(\mathcal{I})^{-1}\mathcal{R} \, & -\mathcal{R}(\mathcal{%
I})^{-1} \\
-(\mathcal{I})^{-1}\mathcal{R} & (\mathcal{I})^{-1}%
\end{array}
\right) \,.
\end{equation}
If the charges satisfy the condition
\begin{equation}
\left\langle \mathcal{G},\mathcal{Q}\right\rangle =-\kappa \,,
\label{eq:kappa}
\end{equation}
the effective action (\ref{eq:S_eff}) can be rewritten as a sum of squares
of first order differential conditions and a boundary term. As in \cite%
{Dall'Agata:2010gj}, setting to zero each of these terms, a system of first
order equations is obtained,
\begin{align}
& 2e^{2\psi }\left( e^{-U}\mathrm{Im}(e^{-i\alpha }\mathcal{V})\right)
^{\prime}+ e^{2(\psi -U)}\Omega \mathcal{M}\mathcal{G}+4e^{-U}(\alpha
^{\prime }+\mathcal{A}_{r})\mathrm{Re}(e^{-i\alpha }\mathcal{V})+\mathcal{Q}%
=0\,,  \nonumber  \label{eq:BPS} \\
& \psi ^{\prime}=2e^{-U}\mathrm{Im}(e^{-i\alpha }\mathcal{L})\,, \\
& \alpha ^{\prime }+\mathcal{A}_{r}=-2e^{-U}\mathrm{Re}(e^{-i\alpha }%
\mathcal{L})\,.  \nonumber
\end{align}
Here $\mathcal{A}_{\mu }=\mathrm{Im}(\partial _{\mu }z^{i}(\partial _{i}%
\mathcal{K}))$ is the connection associated to the K\"{a}hler
transformations and the phase $\alpha $ can be expressed in terms of the
other fields as
\begin{equation}
e^{2i\alpha }=\frac{\mathcal{Z}-ie^{2(\psi -U)}\mathcal{L}}{\bar{\mathcal{Z}}%
+ie^{2(\psi -U)}\bar{\mathcal{L}}}\,.  \label{eq:alpha}
\end{equation}
It is possible to show \cite{Dall'Agata:2010gj} that the supersymmetry
variations of $N=2$ gauged supergravity reproduce the set of equations (\ref%
{eq:BPS}) by requiring the existence of a certain Killing spinor. In this
way, both the equations of motion and the supersymmetry conditions are
satisfied by solutions of (\ref{eq:BPS}), and the resulting configuration
will be 1/4 BPS.

As is the case for many other known solutions \cite%
{Cacciatori:2009iz,Gnecchi:2013mta,Mohaupt:2011aa, Errington:2014bta}, we
shall assume vanishing axions. This is realized by purely imaginary scalars (with $\lambda ^{i}>0$),
\begin{equation}
z^{i}=x^{i}-i\lambda ^{i}\,,\qquad x^{i}=0\,.  \label{eq:vanishing_axions}
\end{equation}
The advantage of this choice will become evident in the next section: for
some values of the FI parameters in $\mathcal{G}$, it indeed simplifies the
equations of motion (\ref{eq:BPS}), setting $\alpha $ to a constant.

\section{\label{nH-STU}A non-homogeneous deformation of the stu model}

In this paper, we will specialize our treatment on the special K\"{a}hler $3$-moduli model
based on the holomorphic prepotential\footnote{Black holes of type IIA Calabi-Yau compactifications
in the presence of perturbative quantum corrections, leading to a prepotential of the form
$F=d_{ijk}X^iX^jX^k/X^0+ic(X^0)^2$ (for some constant $c$), were constructed and studied
in \cite{Bueno:2012jc,Galli:2012pt}.}
\begin{equation}
F=\frac{X^1X^2X^3}{X^0} - \frac A3\frac{\left(X^3\right)^3}{X^0}\,, \label{eq:prepotential-pre}
\end{equation}
where $A$ is an arbitrary real constant. For $A=-1$, the prepotential reads
\begin{equation}
F=\frac{X^{1}X^{2}X^{3}}{X^{0}}+\frac{1}{3}\frac{\left( X^{3}\right) ^{3}}{%
X^{0}}\, ,  \label{eq:prepotential}
\end{equation}
which has been constructed in the context of Type IIA string theory
compactified on Calabi-Yau manifolds in \cite{Louis-1}. In particular,
analyzing string vacua with three complex moduli (section $3.2$ therein),\
different bases for the toric construction of such a model have been
considered; (\ref{eq:prepotential}) corresponds to the basis $\mathbb{F}_{0}$
of \cite{Louis-1}, while other toric constructions determine the same model
in different symplectic frames. The prepotential (\ref{eq:prepotential}) can
also be obtained as $c=0$ limit of the heterotic prepotential appearing in
\cite{Behrndt-1} and the corresponding one-loop prepotential $V_{\text{GS}}$ is
given by considering its $c=0$ limit.

In absence of gauging, the BPS attractor equations for this model have been
discussed in \cite{Behrndt-1}; a solution for a generic supporting black
hole charge configuration was obtained in this context and, as a
consequence, the BPS black hole entropy was determined as a function of the
charges.

A full-fledged, explicit determination of the BPS black hole entropy of the
model based on (\ref{eq:prepotential}) was later given by Shmakova in the
investigation of BPS attractor equations for black holes based on Calabi-Yau
cubic prepotentials \cite{Shmakova}. We report here the expression of the
ungauged BPS black hole entropy, for later convenience:
\begin{equation}
\frac{S_{\text{BH}}}{\pi} = \frac{\sqrt{f\left( \mathcal{Q}\right) }}{3p^{0}},
\label{BH-entropy-1}
\end{equation}%
where%
\begin{equation}
\begin{split}
&f\left( \mathcal{Q}\right) := \\
&2 \left\{ \left( p^{1}p^{2}+\left( p^{3}\right) ^{2}-p^{0}q_{3}\right)
\left[ \left( p^{1}p^{2}+\left( p^{3}\right) ^{2}-p^{0}q_{3}\right)
^{2}+12\left( p^{2}p^{3}-p^{0}q_{1}\right) \left(
p^{1}p^{3}-p^{0}q_{2}\right) \right] \right. \\
& \left. +\left[ \left( p^{1}p^{2}+\left( p^{3}\right)
^{2}-p^{0}q_{3}\right) ^{2}-4\left( p^{2}p^{3}-p^{0}q_{1}\right) \left(
p^{1}p^{3}-p^{0}q_{2}\right) \right] ^{3/2} \right\} \\
&-9\left[ p^{0}\left( p^{0}q_{0}+p^{1}q_{1}+p^{2}q_{2}+p^{3}q_{3}\right)
-2p^{1}p^{2}p^{3}-\frac{2}{3}\left( p^{3}\right) ^{3}\right] ^{2} \, ,
\end{split}
\label{f}
\end{equation}
and the conditions $f\left( \mathcal{Q}\right) >0$ and $p^{0}>0$ define the
BPS-supporting black hole charge vector $\mathcal{Q}$. It is immediate to
check that (\ref{BH-entropy-1}) and (\ref{f}) imply the entropy $S_{\text{BH}}$ to
be homogeneous of degree two in the charges, as it must be in four
dimensions for $0$-branes.

The model (\ref{eq:prepotential-pre}) under consideration, where $A$ has to
be considered a parameter, belongs to the broad class of the so-called very
special K\"{a}hler manifolds, that can be obtained by dimensional reduction
from the vector multiplets' scalar geometries coupled to minimal
supergravity in $D=5$, known as special real manifolds. All the models
originating from this kind of geometry are described, in the so-called
{\it 4D/5D special coordinates'}  symplectic frame (cf.~e.g.~\cite{dWVVP,CFM1}),
by a cubic prepotential of the form
\begin{equation}
F=d_{ijk}\frac{X^{i}X^{j}X^{k}}{X^{0}}\,, \label{eq:d}
\end{equation}
where $d_{ijk}$ is a real and symmetric tensor and the corresponding special
K\"{a}hler space is usually dubbed a $d$-space \cite{dWVVP}. In particular,
the model \eqref{eq:prepotential-pre} is defined by $d_{123}=1/6$ and $d_{333}=-A/3$.

It is worth pointing out that the $d$-space corresponding to
\eqref{eq:prepotential-pre} is neither symmetric nor homogeneous\footnote{After \cite{dWVVP}
and \cite{deWit:1991nm}, homogeneous special K\"{a}hler $d$-spaces, either symmetric or
non-symmetric, have been classified in terms
of the corresponding $d$-tensor, which uniquely determines their geometry.
No homogeneous, non-symmetric, special K\"{a}hler (non-compact, Riemannian)
spaces which are not based on cubic prepotentials (\ref{eq:d}) are known,
although a proof of this fact does not exist, as far as we know.} \cite{Errington:2014bta,deWit:1991nm}.
In particular, it does not fall within the
class of symmetric models examined in \cite{Gnecchi:2013mta}, that are
characterized by a constant tensor\footnote{For some considerations on the completely contravariant
$d$-tensor in generic $d$-spaces (and the corresponding definition of the so-called $E$-tensor for
non-symmetric special K\"{a}hler spaces), cf.~e.g.~\cite{Raju-Quantum-SKG}, and refs.~therein.}
\begin{equation}
\hat{d}^{lmn}=\frac{g^{il}g^{jm}g^{kn}}{(d_{pqr}\lambda ^{p}\lambda
^{q}\lambda ^{r})^{2}}d_{ijk}\,.
\end{equation}%
In fact, it can be easily checked that the prepotential (\ref{eq:prepotential-pre}) implies a non-constant $\hat{d}^{lmn}$.
For this reason, we will henceforth dub the cubic model (\ref{eq:prepotential-pre}) as a \textit{non-homogeneous deformation} of the
homogeneous and symmetric stu model (shortly, nh-stu), to which it reduces\footnote{Consistently,
for $A=0$ the expression (\ref{finite}) below enhances to an $8$-dimensional $U$-duality group,
given by the $\text{SL}(2,\mathbb{R})^{\otimes 3}$
group of the stu model \cite{Duff:1995sm,Behrndt:1996hu,Bellucci:2008sv} (cf.~e.g.~section 8
of \cite{Ruef}).\\
It is here worth pointing out that, however, at the level of the solution
discussed in sections \ref{FI-Gauging} and \ref{Solution} (characterized by
proportionality between $\lambda ^{2}$ and $\lambda ^{3}$), $A=0$ yields the
(axion-free) $\text{st}^2$ model, with some subtleties mentioned at the end of
section \ref{Solution}.} when $A=0$.

\subsection{\label{U-Duality}Electric-magnetic duality algebra}

The vector multiplets' scalar manifold of the nh-stu model is neither symmetric
nor homogeneous; namely, the non-compact Riemannian space endowed with the
special K\"{a}hler geometry specified by the cubic holomorphic prepotential
(\ref{eq:prepotential-pre}) (with non-vanishing $A$) cannot be described as
a coset\footnote{In order for the coset $G/H$ to be non-compact, $H$ must \textit{at least}
be the maximal compact subgroup of $G$. When this is the case, and when both
$G$ and $H$ are reductive Lie groups, the corresponding coset is symmetric.}
$G/H$, where $H$ is a local, compact isotropy group (linearly realized on
the scalar fields, which generally sit in its representations) and $G$ is a
global, non-compact symmetry group (non-linearly realized by the scalar
fields, but linearly realized by the vectors). In theories of Abelian
Maxwell fields, the group $G$ describes the electric-magnetic duality
symmetry, and its non-compactness in presence of scalar fields was firstly
discussed by Gaillard and Zumino in \cite{GZ}.

Linearly realized electric-magnetic duality ($U$-duality\footnote{Here, $U$-duality is referred to as
the ‘continuous' symmetries of \cite{CJ-1}. Their discrete
versions are the $U$-duality non-perturbative string theory symmetries
introduced by Hull and Townsend \cite{HT-1}.}) plays a key role in
Einstein-Maxwell theories coupled to scalar fields in presence of local
supersymmetry, and consequently in their regular solutions, such as the
dyonic black holes discussed in the present paper. Even if the scalar
manifold is not a coset $G/H$, a global $U$-duality symmetry group $G$
always exists, even if it may be non-reductive or also discrete in generic,
(semi-)realistic models of string compactifications.

A general feature of Einstein-Maxwell theories coupled to non-linear sigma
models of scalar fields in four dimensions is the symplectic structure of
the field strength 2-forms and of their duals, which in turn allows to
define the symplectic invariant scalar product specified in footnote \ref{ftn:product}. It results in the
maximal, generally non-symmetric embedding \cite{Dynkin,GZ}
\begin{eqnarray}
G &\subset &\text{Sp}(2n,\mathbb{R}) \\
\mathbf{R} &=&\mathbf{2n}\,,
\end{eqnarray}
where $n$ is the number of vector fields, $\mathbf{2n}$ is the fundamental
representation of $\text{Sp}(2n,\mathbb{R})$ and $\mathbf{R}$ is the representation
of $G$, not necessarily irreducible.\\
Thus, it is interesting to determine the (continuous, Lie component
of the) $U$-duality algebra $\mathfrak{g}_{\text{nh-stu}}$ of the nh-stu model of $N=2$, $D=4$
supergravity. In this case we have $n=4$, since one graviphoton and
three vectors from the vector multiplets are present. We aim to explicitly
find the realization of the maximal, non-symmetric embedding
\begin{equation}
\mathfrak{g}_{\text{nh-stu}}\subset \mathfrak{sp}(8,\mathbb{R})\,. \label{sympl-emb}
\end{equation}
This is worth also in view of the fact that $G_{\text{nh-stu}}$, the Lie group
generated by $\mathfrak{g}_{\text{nh-stu}}$, has not a transitive action on the
non-linear sigma model described by the $N=2$ holomorphic prepotential (\ref{eq:prepotential-pre}).

Since the semiclassical Bekenstein-Hawking entropy in the ungauged theory is
generally invariant under linearly realized global symmetries, $\mathfrak{g}_{\text{nh-stu}}$
can be determined by finding all infinitesimal symplectic
transformations which leave the BPS black hole entropy $S_{\text{BH}}$
(\ref{BH-entropy-1})-(\ref{f}) invariant.\\
Let us choose $A=-1$. From (\ref{BH-entropy-1})-(\ref{f}), the
infinitesimal invariance condition reads
\begin{eqnarray}
\delta S_{\text{BH}} &=&\frac{1}{2S_{\text{BH}}}\delta S_{\text{BH}}^2 \nonumber \\
&=&\frac{1}{2S_{\text{BH}}}\left[ \left( -\frac{2f}{p^{0}}+\frac{\partial f}{\partial p^{0}}\right)
\delta p^0 + \frac{\partial f}{\partial p^{i}}\delta
p^i + \frac{\partial f}{\partial q_{0}}\delta q_{0}+\frac{\partial f}{\partial q_{i}}\delta q_{i}\right] =0\,,
\end{eqnarray}
or equivalently
\begin{equation}
-6f\delta p^{0}+p^{0}\delta f=0,  \label{Eqq}
\end{equation}
where $\delta f=\frac{\partial f}{\partial \mathcal{Q}}\delta\mathcal{Q}$ and
\begin{equation}
\delta \mathcal{Q}=\left( \delta p^{0},\delta p^{i},\delta q_{0},\delta
q_{i}\right) ^{T}=\mathcal{SQ},
\end{equation}
with $\mathcal{S}$ belonging to the symplectic Lie algebra. It is an $8\times 8$ matrix which can be
written in blocks as
\begin{equation}
\mathfrak{sp}\left( 8,\mathbb{R}\right) \ni \mathcal{S}=\left(
\begin{array}{cc}
A & B \\
C & D
\end{array}
\right)\,, \quad A^T=-D\,, \quad B^T=B\,, \quad C^T=C\,,
\end{equation}
where each block is a $4\times 4$ matrix. Thus, $\mathcal{S}$ depends on ten real parameters.

By solving (\ref{Eqq}) for a BPS-supporting configuration with generic
charges $\mathcal{Q}$ satisfying $f\left( \mathcal{Q}\right) >0$ and $p^{0}>0$,
the symplectic embedding of the $U$-duality Lie algebra $\mathfrak{g}_{\text{nh-stu}}$ of the nh-stu
model into $\mathfrak{sp}(8,\mathbb{R})$ is realized by the following four-dimensional, lower triangular
matrix subalgebra (cf.~(\ref{sympl-emb}); $a,b,c\in \mathbb{R}$, $\phi \in \mathbb{R}_{0}^{+}$)
\begin{equation}
\mathcal{S}_{\text{nh-stu}}(a,b,c,\phi) =\left(
\begin{array}{cccccccc}
-3\phi  & 0 & 0 & 0 & 0 & 0 & 0 & 0 \\
a & -\phi  & 0 & 0 & 0 & 0 & 0 & 0 \\
b & 0 & -\phi  & 0 & 0 & 0 & 0 & 0 \\
c & 0 & 0 & -\phi  & 0 & 0 & 0 & 0 \\
0 & 0 & 0 & 0 & 3\phi  & -a & -b & -c \\
0 & 0 & c & b & 0 & \phi  & 0 & 0 \\
0 & c & 0 & a & 0 & 0 & \phi  & 0 \\
0 & b & a & 2c & 0 & 0 & 0 & \phi
\end{array}
\right) \in \mathfrak{g}_{\text{nh-stu}}\subset \mathfrak{sp}(8,\mathbb{R})\,.  \label{S-call}
\end{equation}
For a generic $A$, this can be generalized as follows:
\begin{equation}
\mathcal{S}_{\text{nh-stu}}(a,b,c,\phi;A) =\left(
\begin{array}{cccccccc}
-3\phi  & 0 & 0 & 0 & 0 & 0 & 0 & 0 \\
a & -\phi  & 0 & 0 & 0 & 0 & 0 & 0 \\
b & 0 & -\phi  & 0 & 0 & 0 & 0 & 0 \\
c & 0 & 0 & -\phi  & 0 & 0 & 0 & 0 \\
0 & 0 & 0 & 0 & 3\phi  & -a & -b & -c \\
0 & 0 & c & b & 0 & \phi  & 0 & 0 \\
0 & c & 0 & a & 0 & 0 & \phi  & 0 \\
0 & b & a & -2Ac & 0 & 0 & 0 & \phi
\end{array}
\right) \in \mathfrak{g}_{\text{nh-stu}}\subset \mathfrak{sp}(8,\mathbb{R})\,. \label{S-call-1}
\end{equation}
It can be noticed that (\ref{S-call-1}) (which reduces to (\ref{S-call}) for
$A=-1$) determines a maximal Abelian subalgebra of $\mathfrak{sp}(8,\mathbb{R})$, whose four
generators commute. Moreover, the
generators corresponding to $a,b,c$ in (\ref{S-call}) span an axionic
Peccei-Quinn translational three-dimensional algebra, nilpotent of
degree four. Indeed, the part of (\ref{S-call}) generated by $a,b,c$ can be
recast in the following generic, $d$-parametrized form \cite{Andrianopoli:2002mf}
\begin{equation}
\mathcal{S}=\begin{pmatrix} 0 & 0 & 0 & 0 \\ a^{j} & 0 & 0 & 0 \\ 0 & 0 & 0
& -a^{i} \\ 0\, & d_{a,ij}\, & 0\, & 0\,\end{pmatrix}\subset \mathfrak{sp}(2n,\mathbb{R})\,,  \label{S}
\end{equation}
where ($i=1,...,n-1$)
\begin{equation}
\begin{split}
&
d_{a,ij}:=d_{ijk}a^{k}\,, \qquad d_{a,i}:=d_{ijk}a^{j}a^{k}\,, \qquad d_{a}:=d_{ijk}a^{i}a^{j}a^{k}\,, \\
& a^{1}:=6a\,, \qquad a^{2}:=6b\,, \qquad a^{3}:=6c\,.
\end{split}
\label{dcontractions}
\end{equation}
$\mathcal{S}$ in (\ref{S}) can be easily checked to be nilpotent of degree
four\footnote{$\mathbb{I}_{d}$ denotes the $d\times d$ identity matrix
throughout.},
\begin{equation}
\mathcal{S}^{4}=0\Rightarrow \exp \left( \mathcal{S}\right) =\mathbb{I}_{2n}+
\mathcal{S}+\frac{1}{2}\mathcal{S}^{2}+\frac{1}{3!}\mathcal{S}^{3},
\end{equation}
yielding, at group level \cite{Bellucci:2010aq, d-geom},
\begin{equation}
\exp \left( \mathcal{S}\right) =\begin{pmatrix} 1 & 0 & 0 & 0 \\ a^{j} &
\mathbb{I}_{n-1} & 0 & 0 \\ \,-\frac{1}{6}d_a \, & \,-\frac{1}{2}d_{a,i} \,
&\, 1 \, &\,-a^{i}\, \\ \frac{1}{2}d_{a,j} & d_{a,ij} & 0 &
\mathbb{I}_{n-1}\end{pmatrix}\subset \text{Sp}(2n,\mathbb{R})\,.
\end{equation}
Such an Abelian $\left( n-1\right) $--dimensional global symmetry
algebra/group, as discussed in \cite{d-geom} (see also refs.~therein, in
particular \cite{Strom}), characterizes \textit{every} model of $D=4$
supergravity based on a cubic scalar geometry, even not of special K\"{a}hler type
(i.e.~the scalar geometries of $N=4$, $6$ and $8$
supergravity theories, dubbed ‘generalized $d$-geometries'
in \cite{d-geom}): the representation of axions in $D=4$ is \textit{always}
nilpotent of degree four.\\
Besides the $\left( n-1\right) $-dimensional axionic Peccei-Quinn
translational algebra, the universal sector of the electric-magnetic duality
algebra of every (generalized) $d$-geometry (also cf. \cite{Bellucci:2010zd}) is given by the $2n\times 2n$
generalization of the $\phi $-parametrized part of (\ref{S-call-1}), where $%
\phi $ can be thus regarded as the Kaluza-Klein radius/real dilaton of the
Kaluza-Klein (KK) $\mathfrak{so}_{\text{KK}}(1,1)$,
\begin{equation}
K(\phi) =\left(
\begin{array}{cccc}
-3\phi  & 0 & 0 & 0 \\
0 & -\phi \mathbb{I}_{n-1} & 0 & 0 \\
0 & 0 & 3\phi  & 0 \\
0 & 0 & 0 & \phi \mathbb{I}_{n-1}
\end{array}
\right) \in \mathfrak{so}_{\text{KK}}(1,1)\subset \mathfrak{sp}(2n,\mathbb{R})\,.
\end{equation}
Therefore, the $2n\times 2n$ matrix realization of the universal sector of
the global electric-magnetic duality symmetry of an Einstein-Maxwell theory
whose scalar manifold is endowed with a ‘generalized $d$-geometry' can be written at the Lie
algebra level as \cite{Bellucci:2010aq, d-geom}
\begin{equation}
\mathcal{S}(a) +K(\phi) =\left(
\begin{array}{cccc}
-3\phi  & 0 & 0 & 0 \\
a^{j} & -\phi \delta _{i}^{j} & 0 & 0 \\
0 & 0 & 3\phi  & -a^{i} \\
0 & d_{a,ij} & 0 & \phi \delta _{j}^{i}
\end{array}
\right) \subset \mathfrak{sp}(2n,\mathbb{R})\,,
\end{equation}
and at the Lie group level as \cite{Bellucci:2010aq, d-geom}
\begin{equation}
\exp\!\left(\mathcal{S}(a)\right)\exp\!\left( K(\phi)\right) =\left(
\begin{array}{cccc}
e^{-3\phi } & 0 & 0 & 0 \\
a^{j} & e^{-\phi }\delta _{i}^{j} & 0 & 0 \\
-\frac{1}{6}d_{a}\, & -\frac{1}{2}d_{a,i}\, & e^{3\phi }\, & -a^{i}\, \\
\frac{1}{2}d_{a,j} & d_{a,ij} & 0 & e^{\phi }\delta _{j}^{i}
\end{array}
\right) \subset \text{Sp}(2n,\mathbb{R})\,. \label{finite}
\end{equation}
When considering $N=2$, $D=4$ theories, this result for special K\"{a}hler $d$-geometries was known
after\footnote{In \cite{d-geom}, (\ref{finite}) was shown also to pertain to the universal
sector of axionic and KK coordinates in the scalar manifolds of $D=4$
theories based on ‘generalized $d$-geometries' (for
non-homogeneous $N=2$ very special K\"{a}hler geometries, the same
parametrization provides a general description of the generic element of the
\textit{flat} symplectic bundle over the vector multiplets' scalar manifold
\cite{Strom,d-geom}).} \cite{dWVVP}.\\
Thus, in this sense, one can conclude that the nh-stu model has the \textit{smallest possible}
electric-magnetic duality algebra, consistent with its
cubic nature (and thus with its upliftability to $N=1$, $D=5$ supergravity).

\subsection{\label{Axion-Free-Geom}Axion-free geometry}

As stated above, in the present investigation we consider only the
axion-free case, thus parametrising the purely imaginary scalar fields as $%
z^{i}=-i\lambda ^{i}$, with $\lambda ^{i}$ real and positive ($i=1,2,3$); we
are also choosing the projective coordinates as
\begin{equation}
\frac{X^{1}}{X^{0}}=-i\lambda ^{1}\,,\qquad \frac{X^{2}}{X^{0}}=-i\lambda
^{2}\,,\qquad \frac{X^{3}}{X^{0}}=-i\lambda ^{3}\,.
\end{equation}
Thus, the symplectic sections (\ref{eq:section}) become ($\Lambda =0,1,2,3$)
\begin{equation}
\begin{split}
& L^{\Lambda }=e^{\mathcal{K}/2}\left( 1,-i\lambda ^{1},-i\lambda
^{2},-i\lambda ^{3}\right) ^{T}\,, \\
& M_{\Lambda }=e^{\mathcal{K}/2}\left( -i\left( \lambda ^{1}\lambda
^{2}\lambda ^{3}-\frac{A}{3}(\lambda ^{3})\mbox{}^{3}\right) ,-\lambda
^{2}\lambda ^{3},-\lambda ^{1}\lambda ^{3},-\lambda ^{1}\lambda
^{2}+A(\lambda ^{3})\mbox{}^{2}\right) ^{T}\,,
\end{split}
\end{equation}
while the K\"{a}hler potential reads
\begin{equation}
e^{-\mathcal{K}}=8\left( \lambda ^{1}\lambda ^{2}\lambda ^{3}-\frac{A}{3}
(\lambda ^{3})\mbox{}^{3}\right)\,.
\end{equation}
For vanishing axions, the special K\"{a}hler metric takes the form
\begin{equation}
g_{i\bar{\jmath}}=\frac{1}{4\left( \lambda ^{1}\lambda ^{2}\lambda ^{3}-
\frac{A}{3}(\lambda ^{3})\mbox{}^{3}\right) ^{2}}\left(
\begin{array}{ccc}
(\lambda ^{2})\mbox{}^{2}(\lambda ^{3})\mbox{}^{2} & \frac{A}{3}(\lambda
^{3})\mbox{}^{4} & -\frac{2}{3}A\lambda ^{2}(\lambda ^{3})\mbox{}^{3} \\
&  &  \\
\frac{A}{3}(\lambda ^{3})\mbox{}^{4}\; & (\lambda ^{1})\mbox{}^{2}(\lambda
^{3})\mbox{}^{2} & -\frac{2}{3}A\lambda ^{1}(\lambda ^{3})\mbox{}^{3} \\
&  &  \\
-\frac{2}{3}A\lambda ^{2}(\lambda ^{3})\mbox{}^{3}\; & -\frac{2}{3}A\lambda
^{1}(\lambda ^{3})\mbox{}^{3} & \;\;(\lambda ^{1})\mbox{}^{2}(\lambda ^{2})%
\mbox{}^{2}+\frac{A^{2}}{3}(\lambda ^{3})\mbox{}^{4}%
\end{array}%
\right) \,.  \label{eq:Kaehler_metric}
\end{equation}
The symplectic matrix $\mathcal{N}_{\Lambda \Sigma }$ has, in the axion-free
case under consideration, vanishing real part $\mathcal{R}_{\Lambda \Sigma }$, while
$\mathcal{I}_{\Lambda \Sigma }$ is given by
\begin{equation}
\mathcal{I}_{\Lambda \Sigma }=-\frac{1}{8}e^{-\mathcal{K}}\left(
\begin{array}{cc}
1 & 0 \\
0 & 4g_{i\bar{\jmath}}
\end{array}
\right)\,,
\end{equation}
which is thus consistently negative definite.

\section{\label{FI-Gauging}Dyonic Fayet-Iliopoulos gaugings and near-horizon analysis}

To proceed further, we shall assume a specific form for the FI parameters $\mathcal{G}$. The choice
\begin{equation}
\mathcal{G}^{T}=(0,g^{1},g^{2},g^{3},g_{0},0,0,0)^{T}\,, \label{eq:G}
\end{equation}
together with the vanishing axion condition (\ref{eq:vanishing_axions}),
fixes the phase $\alpha $ in (\ref{eq:alpha}) to the constant value\footnote{Another possible choice
yielding the same constant value for $\alpha $ is $\mathcal{G}^{T}=(g^0,0,0,0,0,g_1,g_2,g_3)^T$,
which would in turn
require $\mathcal{Q}$ to assume the (magnetic) form $\mathcal{Q}^T=(0,p^1,p^2,p^3,q_0,0,0,0)^T$.} $\alpha =\pm \pi /2$. This
can be checked from the explicit expressions of the symplectic invariants $%
\mathcal{Z}$ and $\mathcal{L}$,
\begin{equation}
\begin{split}
& \mathcal{Z}=ie^{\mathcal{K}/2}\left( p^{0}\left( \lambda ^{1}\lambda
^{2}\lambda ^{3}-\frac{A}{3}\lambda ^{3}\right) -q_{1}\lambda
^{1}-q_{2}\lambda ^{2}-q_{3}\lambda ^{3}\right) \,, \\
& \mathcal{L}=e^{\mathcal{K}/2}\left( g_{0}+g^{1}\lambda ^{2}\lambda
^{3}+g^{2}\lambda ^{1}\lambda ^{3}+g^{3}(\lambda ^{1}\lambda ^{2}-A(\lambda
^{3})\mbox{}^{2})\right) \,.
\end{split}
\end{equation}
As can be inferred from the BPS equations (\ref{eq:BPS}), the choice (\ref{eq:G}) requires some charges
to vanish, so that the vector $\mathcal{Q}$ takes the form
\begin{equation}
\mathcal{Q}^{T}=(p^{0},0,0,0,0,q_{1},q_{2},q_{3})^{T}\,. \label{eq:Q}
\end{equation}
With the choice \eqref{eq:G}, the scalar potential (\ref{eq:scalar_pot})
becomes
\begin{eqnarray}
V_{g} &=&-g^{2}g^{3}\lambda ^{1}-g^{1}g^{3}\lambda ^{2}-\left(
g^{1}g^{2}-A(g^{3})^{2}\right) \lambda ^{3}  \nonumber \\
&&-\frac{g_{0}}{\lambda ^{1}\lambda ^{2}\lambda ^{3}-\frac{A}{3}(\lambda
^{3})\mbox{}^{3}}\left( g^{2}\lambda ^{1}\lambda ^{3}+g^{1}\lambda
^{2}\lambda ^{3}+g^{3}\left( \lambda ^{1}\lambda ^{2}-A(\lambda ^{3})\mbox{}%
^{2}\right)\right)\,, \label{pot-axionfree}
\end{eqnarray}
which matches the known expression for the stu model \cite{CFM1,Bellucci:2008sv,Cacciatori:2009iz}
for $A=0$. In what follows we shall assume that all gauge coupling constants $g_0,g^i$ are positive.
Then the potential \eqref{pot-axionfree} has two critical points, namely one for
\begin{equation}
\lambda^1 = \frac{g^1}{g^3}\lambda^3\,, \qquad \lambda^2 = \frac{g^2}{g^3}\lambda^3\,, \qquad
\lambda^3 = \sqrt{\frac{g_0g^3}{g^1g^2 - \frac A3(g^3)^2}}\,, \label{usual-crit-pt}
\end{equation}
and the other for
\begin{equation}
\lambda^1 = \frac{g^1}{g^3}\lambda^3\,, \qquad \lambda^2 = -\frac1{g^1g^3}\left(g^1g^2 -
\frac23 A(g^3)^2\right)\lambda^3\,, \qquad \lambda^3 = \sqrt{\frac{g_0g^3}{g^1g^2 -
\frac A3(g^3)^2}}\,. \label{new-crit-pt}
\end{equation}
The first has $V_g=-3\ell^{-2}$, and the second $V_g=-\ell^{-2}$, with $\ell$ defined in \eqref{ell},
so both correspond to AdS vacua. One easily shows that \eqref{usual-crit-pt} is also a critical point of
the superpotential \eqref{eq:L}, while \eqref{new-crit-pt} is not. The vacuum \eqref{usual-crit-pt}
is thus supersymmetric, whereas \eqref{new-crit-pt} breaks supersymmetry. Moreover, reality and
positivity of the scalars $\lambda^i$ implies that the second vacuum exists only in the range
\begin{equation}
\frac32\frac{g^1g^2}{(g^3)^2} < A < 3\frac{g^1g^2}{(g^3)^2}\,,
\end{equation}
in particular it is not present for zero deformation parameter $A$.

Owing to the constancy of $\alpha$, the equations of motion (\ref{eq:BPS})
boil down to
\begin{equation}
\begin{split}  \label{eq:to_solve}
& 2 e^{2\psi}\left(e^{-U}\mathrm{Re}\mathcal{V}\right)^{\prime}+e^{2(%
\psi-U)}\Omega\mathcal{M}\mathcal{G} + \mathcal{Q} = 0\,, \\
& (e^\psi)^{\prime}=2e^{\psi-U } \mathrm{Re}\mathcal{L}\,.
\end{split}
\end{equation}
The near-horizon geometry is required to be $\text{AdS}_2\times\Sigma $, i.e.,
the metric functions in \eqref{eq:metric} should take the form
\begin{equation}
e^{U}=\frac{r}{R_{1}}\,,\qquad e^{\psi }=r\,\frac{R_{2}}{R_{1}}\,, \label{eq:Upsi-nearhor}
\end{equation}
while the scalar fields $z^{i}(r)=-i\lambda ^{i}(r)$ assume a constant value on the horizon.
Under this assumption, the BPS equations \eqref{eq:to_solve} simplify to
\begin{equation}
\begin{split}
& \mathcal{Q}+R_{2}^{2}\Omega \mathcal{M}\mathcal{G}=-4\mathrm{Im}\left(
\overline{\mathcal{Z}}\mathcal{V}\right) \,, \\
& \mathcal{Z}=i\,\frac{R_{2}^{2}}{2R_{1}}\,.
\end{split}
\label{eq:nearhor}
\end{equation}
In addition, one has to impose the constraint \eqref{eq:kappa}.\\
Following the procedure described in \cite{Halmagyi:2013qoa}\footnote{The
equations \eqref{eq:attr-nh-stu} are based on \cite{Halmagyi:2013qoa}, with some misprints
corrected.}, the BPS equations in the near-horizon limit
\eqref{eq:nearhor} provide a set of equations for the variables $\{R_1,R_2,\lambda^i\} $ as functions
of the gaugings $g_0,g^i$ and the charges $p^0,q_i$.\newline
In particular, since $R_2$ is directly related to the black hole entropy $S$, this yields an expression
for $S$ in terms of the gaugings and charges. In the model described above, the attractor equations
\eqref{eq:nearhor} are implicitly solved by
\begin{equation} \label{eq:attr-nh-stu}
\begin{split}
& R_{2}^{4}\,d_{g,i}+\frac{1}{3}\left( \kappa +\frac{1}{2}\right) p^{0}q_{i}=%
\frac{1}{36}\left( d_{\lambda }^{-1}\right)^{ij}q_{j}\,q_{i}-\frac{1}{4}%
\left( p^{0}\right) ^{2}d_{\lambda ,i}\,, \\
& \lambda ^{i}\left( 1-\frac{\kappa }{2}\right) =\frac{\kappa }{p^{0}}\left(
-R_{2}^{2}\,g^{i}+\frac{1}{6}\left( d_{\lambda }^{-1}\right)
^{ij}q_{j}\right) \,, \\
& \frac{R_{2}^{2}}{R_{1}}=\left( p^{0}e^{-\frac{\mathcal{K}}{2}}\left(
\kappa -\frac{3}{4}\right) -2\,e^{\frac{\mathcal{K}}{2}}\lambda
^{j}q_{j}\right) \,, \\
& R_{2}^{6}\,d_{g}+\frac{1}{2}R_{2}^{2}\,p^{0}\left( p^{0}g_{0}+\kappa
g^{i}q_{i}\right) =\frac{1}{216}\left( d_{\lambda }^{-1}\right) ^{k}\left(
d_{\lambda }^{-1}\right) ^{ij}q_{i}\,q_{j}\,q_{k} \\
& \hspace{0.95cm}+\frac{1}{64}p^{0}q_{i}\,q_{j}\left( \left( d_{\lambda
}^{-1}\right) ^{j}\lambda ^{i}+2\left( d_{\lambda }^{-1}\right) ^{ij}\right)
+\frac{1}{8}\left( p^{0}\right) ^{2}\left( \lambda ^{i}q_{i}+p^{0}d_{\lambda
}\right) \,,
\end{split}
\end{equation}
where the contractions of the tensor $d_{ijk}$ are defined as in (\ref{dcontractions}). Note that the
non-homogeneity enters through $(d_{\lambda }^{-1})^{ij}$, that depends on the special K\"ahler
metric, since
\begin{displaymath}
g^{ij} = -\frac23 d_\lambda(d_{\lambda }^{-1})^{ij} + 2\lambda^i\lambda^j\,,
\end{displaymath}
cf.~eq.~(A.6) of \cite{Halmagyi:2013qoa}.\\
An explicit solution to \eqref{eq:attr-nh-stu} cannot be obtained by applying the analysis developed
in \cite{Halmagyi:2013qoa} for the case of symmetric special K\"ahler manifolds, because the model
under consideration is neither symmetric nor homogeneous.

\section{\label{Solution}The full black hole solution}

The present section is devoted to the presentation of an exact black hole
solution for the nh-stu model introduced in section \ref{nH-STU}. In order to
simplify the BPS equations \eqref{eq:to_solve}, we introduce the functions\footnote{A common choice
for the functions $H_{i}$ is to make them coincide with the
components of the symplectic sections. For the present situation, we
preferred to choose $H_{3}$ in a different way, in order to simplify the
structure of the equations.}
\begin{equation}
\begin{split}
H^{0}& =\frac{e^{-U}}{\sqrt{2}}\left( \lambda ^{1}\lambda ^{2}\lambda ^{3}-%
\frac{A}{3}(\lambda ^{3})\mbox{}^{3}\right) ^{\!-\frac{1}{2}}, \\
H_{1}& =\lambda ^{2}\lambda ^{3}H^{0}\,,\qquad H_{2}=\lambda ^{1}\lambda
^{3}H^{0}\,,\qquad H_{3}=(\lambda ^{3})\mbox{}^{2}H^{0}\,.  \label{eq:def_H}
\end{split}
\end{equation}
In terms of the latter, the equations \eqref{eq:to_solve} become
\begin{equation}
\begin{split}
& (H^{0})^{\prime }+2g_{0}(H^{0})^{2}=-e^{-2\psi }p^{0}\,, \\
& H_{1}^{\prime 1}H_{1}^{2}+\frac{2}{3}Ag^{2}H_{3}^{2}-\frac{4}{3}%
Ag^{3}H_{1}H_{3}=e^{-2\psi }q_{1}\,, \\
& H_{2}^{\prime 2}H_{2}^{2}+\frac{2}{3}Ag^{1}H_{3}^{2}-\frac{4}{3}%
Ag^{3}H_{2}H_{3}=e^{-2\psi }q_{2}\,, \\
& H_{3}^{\prime }+2H_{3}(g^{1}H_{1}+g^{2}H_{2})-2g^{3}\left( H_{1}H_{2}+%
\frac{A}{3}H_{3}^{2}\right) = \\
& \hspace*{0.5cm}=e^{-2\psi }\frac{H_{3}}{H_{1}H_{2}+AH_{3}^{2}}%
(q_{1}H_{2}+q_{2}H_{1}-q_{3}H_{3})\,, \\
& \psi ^{\prime }=g_{0}H^{0}+g^{1}H_{1}+g^{2}H_{2}+g^{3}\left( \frac{%
H_{1}H_{2}}{H_{3}}-AH_{3}\right) \,.  \label{eq:eq_H}
\end{split}
\end{equation}
A remarkable feature of the nh-stu model is that, contrary to e.g.~the case considered
in \cite{Cacciatori:2009iz}, the equations \eqref{eq:eq_H} cannot be decoupled, due to the nondiagonal
terms in the metric (\ref{eq:Kaehler_metric}).
Following the strategy of \cite{Cacciatori:2009iz}, we use the Ansatz
\begin{equation}
\begin{split}
& \psi =\log \left( a\,r^{2}+c\right) \,, \\
& H^{0}=e^{-\psi }\left( \alpha ^{0}r+\beta ^{0}\right) \,, \\
& H_{i}=e^{-\psi }\left( \alpha _{i}r+\beta _{i}\right) ,\qquad i=1,2,3\,.
\label{eq:Ansatz_H}
\end{split}
\end{equation}
The solution for the fields is then expressed in terms of the functions $H^0,H_i$ by inverting the
relations (\ref{eq:def_H}). This yields
\begin{equation}
e^{2U}=\frac{1}{2}\left( \frac{H_{3}}{H^{0}}\right)^{\frac12}\left(
H_{1}H_{2}-\frac{A}{3}H_{3}^{2}\right)^{-1}\,,
\end{equation}
and
\begin{equation}
\lambda^1 = H_{2}\left( H_{3}H^{0}\right) ^{-\frac{1}{2}}\,, \qquad
\lambda^2 = H_{1}\left( H_{3}H^{0}\right) ^{-\frac{1}{2}}\,, \qquad
\lambda^3 = \left(\frac{H_{3}}{H^{0}}\right) ^{\frac{1}{2}}\,,
\label{eq:lambda_H}
\end{equation}
for the warp factor and the scalars respectively.
By means of the Ansatz (\ref{eq:Ansatz_H}), the differential equations (\ref{eq:to_solve}) boil down
to a system of algebraic conditions on the
parameters and the charges characterizing the solution,
i.e.,~$\{\alpha^0,\alpha_i,\beta^0,\beta_i,a,c,p^0,q_i\}$. The set
of equations obtained in this way reduces, after some algebraic
manipulations, to
\begin{equation}
\begin{split}
\alpha ^{0}=& \,\frac{a}{2g_{0}}\,,\qquad \alpha _{1}=\,\frac{g^{2}}{g^{3}}%
\,\alpha _{3}\,,\qquad \alpha _{2}=\,\frac{g^{1}}{g^{3}}\,\alpha _{3}\,, \qquad
\alpha _{3} = \frac{a\,g^{3}}{2\left( g^{1}g^{2}-\frac{A}{3}\left(
g^{3}\right) ^{2}\right) }\,, \\
\beta _{1}=& \,\frac{g^{2}}{g^{3}}\,\beta _{3}\,,\qquad \beta _{2}=\,-\frac{1%
}{2}\,\beta _{3}\left( \frac{g^{1}}{g^{3}}-A\frac{g^{3}}{g^{2}}\right) -%
\frac{1}{2}\,\beta ^{0}\,\frac{g_{0}}{g^{2}}\,, \\
q_{1}=& \,2\,\beta _{3}^{2}\,\frac{g^{2}}{\left( g^{3}\right) ^{2}}\left(
g^{1}g^{2}-\frac{A}{3}\left( g^{3}\right) ^{2}\right) +g^{2}\,\frac{ac}{%
2\left( g^{1}g^{2}-\frac{A}{3}\left( g^{3}\right) ^{2}\right) }\,, \\
q_{2}=& \,\frac{1}{2g^{2}}\left( \beta ^{0}g_{0}+\beta _{3}\,\frac{g^{1}g^{2}%
}{g^{3}}\right) ^{2}+g^{1}\,\frac{ac}{2\left( g^{1}g^{2}-\frac{A}{3}\left(
g^{3}\right) ^{2}\right) } \\
& +\frac{A}{3}\,\beta _{3}\,\frac{g^{3}}{g^{2}}\left( \beta _{3}\,\frac{%
g^{1}g^{2}}{g^{3}}-\beta ^{0}g_{0}-\frac{A}2\beta_3 g^3\right)\,, \\
q_{3}=& \,\frac{g^{2}}{g^{3}}\,q_{2}-A\,\frac{g^{3}}{g^{2}}\,q_{1}\,, \qquad
p^0 = -\frac{ac}{2g_{0}}-2g_{0}\left( \beta ^{0}\right) ^{2}\,. \\
\label{eq:solution}
\end{split}
\end{equation}
The solution for the scalars is obtained by plugging the parameters written in (\ref{eq:solution}) into
the expressions (\ref{eq:lambda_H}). In this way, the scalars assume the explicit form
\begin{equation} \label{expl-form-scalars}
\begin{split}
& \lambda ^{1}=\frac{a\,\frac{g^{1}}{g^{3}}\left( \lambda _{\infty
}^{3}\right) ^{2}r-g_{0}\,\beta _{3}\left( \frac{g^{1}}{g^{3}}-A\,\frac{g^{3}%
}{g^{2}}\right) -\beta ^{0}\,\frac{g_{0}^{2}}{g^{2}}}{\sqrt{\left(
2g_{0}\,\beta ^{0}+a\,r\right) \left( 2g_{0}\,\beta _{3}+a\,r\left( \lambda
_{\infty }^{3}\right) ^{2}\right) }}\,, \\
& \lambda ^{2}=\frac{g^{2}}{g^{3}}\lambda ^{3}\,, \qquad
\lambda ^{3}=\lambda _{\infty }^{3}\sqrt{\frac{ar+\,\frac{2\,g_{0}}{%
(\lambda _{\infty }^{3})\mbox{}^{2}}\,\beta _{3}}{ar+2g_0\beta^0}}\,,
\end{split}
\end{equation}
where $\lambda _{\infty }^{3}$ is the asymptotic value of $\lambda ^{3}$,
\begin{equation}
\lambda_{\infty }^3 = \sqrt{\frac{g_0 g^3}{g^1 g^2 - \frac A3 (g^3)^2}}\,.
\end{equation}
The warp factor in the metric reads
\begin{equation}
\begin{split}
& \mathrm{e}^{2U}=\frac{2g_0 g^{3}(ar^2+c)\mbox{}^{2}}{\lambda
_{\infty }^{3}\left( ar-g_0\beta^0 - \frac{\,g_{0}}{(\lambda_{\infty}^3)^2}\beta_3\right)
\sqrt{\left( ar+2g_0\beta^0
\right) \left( ar+\frac{2 g_0}{(\lambda _{\infty }^{3})^2}\beta_3\right)}}\,.
\end{split}
\label{eq:e2U}
\end{equation}
This solution represents a black hole, with a horizon at the largest zero of $e^{2U}$, i.e., at
$r_{\text h}=\sqrt{-c/a}$, where we assumed $a>0$ and $c<0$. The curvature invariants diverge
where the angular component of the metric $e^{2\psi-2U}$ vanishes.
Note that all the scalar fields $\lambda _{i}$ should be well-defined and
positive outside the horizon. Moreover, we still have to impose the condition (\ref{eq:kappa}), i.e.,
\begin{equation}
g_0 p^0 - g^i q_i = -\kappa \label{cond-Dirac}
\end{equation}
on the solution (\ref{eq:solution}).
We checked that these requirements are compatible with any of the three
possible choices for $\kappa =0\,,\pm 1$, i.e., the horizon
topology can be either spherical, flat or hyperbolic.

The Dirac quantization condition \eqref{cond-Dirac} fixes one of the four parameters
$\{a,c,\beta^0,\beta_3\}$ that determine the solution, for example $c$ . Furthermore, one easily sees
that the solution enjoys the scaling symmetry
\begin{equation}
(t,r,\theta,\phi,a,c,\beta^0,\beta_3,\kappa) \mapsto (t/s,sr,\theta,\phi,a/s,sc,\beta^0,\beta_3,\kappa)\,,
\qquad s\in\mathbb{R}\,,
\end{equation}
that can be used to set $a=1$ without loss of generality. Consequently, there are only two
physical parameters left, on which the solution depends.\\
Notice that the solution (\ref{eq:solution}) is characterized by the proportionality
between the scalars $\lambda_2$ and $\lambda_3$, as is evident from \eqref{expl-form-scalars}.
However, it is worth stressing that this fact does not trivialize our results,
since the locus $\lambda^2=\frac{g^2}{g^3}\lambda^3$ in the scalar manifold does not yield a
consistent two-moduli truncation for the model (\ref{eq:prepotential-pre}). In other
words, the K\"{a}hler geometry that can be derived from the truncated model
$F\left( X^{1},X^{2},X^{3})\right\vert _{\lambda ^{2}\propto \lambda^3}$
is not equivalent to the two-dimensional one characterized by the prepotential
\begin{equation}
F=\frac{\tilde{X}^{1}\left( X^{3}\right)^2}{X^0}\,, \qquad \mbox{with}
\qquad \tilde{X}^{1}=X^{1}-\frac{A}{3}\,X^{3}\,,  \label{eq:troncato}
\end{equation}
which is homogeneous and symmetric (the so-called $\text{st}^2$ model, cf.~e.g.~\cite{Bellucci:2007zi}
and refs.~there\-in). This difference is evident, for example, in terms of the K\"{a}hler metric.
In fact one has
\begin{equation}
g_{ij}^{(3)}d\lambda^i d\lambda^j\rvert_{\lambda_2\propto\lambda_3} \neq g_{MN}^{(2)}
d\lambda^M d\lambda^N\,, \qquad i,j=1,2,3\,, \qquad M,N=1,2\,,
\end{equation}
where the left-hand side is the line element obtained with the metric (\ref{eq:Kaehler_metric}) when the condition $\lambda_2\propto\lambda_3$ is imposed, while the right-hand side describes the geometry associated to the prepotential (\ref{eq:troncato}).

We conclude this section with a comment on the behaviour of the
solution for $A=0$. Due to the particular definition of $H_3$ we have chosen (with respect to
the more common one used for example in \cite{Dall'Agata:2010gj,Cacciatori:2009iz,Gnecchi:2013mta}), setting $A=0$ and $\lambda^2=\frac{g^2}{g^3}\lambda^3$ is not sufficient to match
exactly the stu black hole solution with two independent parameters, known as $\text{st}^2$ solution,
that can be derived from \cite{Cacciatori:2009iz}. However, the parameters in (\ref{eq:Ansatz_H}) can
be redefined as
\begin{equation}
\alpha_3^{\prime} = \frac{\alpha_1\alpha_2}{\alpha_3} - \frac A3\alpha_3\,, \qquad
\beta_3^{\prime} = \frac{\beta_1\beta_2}{\beta_3} - \frac A3\beta_3\,,
\end{equation}
in terms of which the solution (\ref{eq:solution}) matches explicitly the
known one when $A=0$. This redefinition of the parameters is a way to
recover the choice for the functions that is usually made when solving the
BPS equations (\ref{eq:BPS}), whose analogue for the present case is
\begin{equation}
H_3^{\prime} = \left(\lambda^1\lambda^2 - A(\lambda^3)^2\right)H^0\,, \quad \text{or}
\quad H_3^{\prime} = e^{-\psi}(\alpha_3^{\prime}r + \beta_3^{\prime})\,.
\end{equation}

\section{\label{Phys-Discussion}Physical discussion}

In this section, we discuss some properties of our solution, like near-horizon limit, entropy or
area-product formula.

In the asymptotic limit $r\rightarrow\infty$, the metric (\ref{eq:e2U}) becomes $\mathrm{AdS}_4$, i.e.,
at leading order one has
\begin{equation}
\mathrm{d}s^2 \to -\frac{r^2}{\ell^2}\mathrm{d}t^2 + \ell^2\frac{\mathrm{d}r^2}{r^2} +
r^2\mathrm{d}\Omega_{\kappa}^2\,,
\end{equation}
where we defined the asymptotic $\text{AdS}_4$ curvature radius $\ell$ by
\begin{equation}
\ell^2 = \frac{\lambda_\infty^3}{2g_0g^3}\,, \label{ell}
\end{equation}
and rescaled the coordinates according to $t\to\ell t$, $r\to r/\ell$. Notice that the asymptotic value
of the cosmological constant is
\begin{equation}
\Lambda = -\frac3{\ell^2} = -\frac{6g_0g^3}{\lambda_\infty^3}\,.
\end{equation}
On the other hand, when $r$ approaches the horizon $r_{\text h}$, the functions $U$ and $\psi$ assume,
after shifting $r\to r+r_{\text h}$, the form \eqref{eq:Upsi-nearhor}, with $R_1$ and $R_2$ given by
\begin{equation}
R_1^2 = -\frac{\lambda_\infty^3 f(r_{\text h})}{8g_0g^3c}\,, \qquad R_2^2 =
\frac{\lambda_\infty^3 f(r_{\text h})}{2g_0g^3}\,,
\end{equation}
where
\begin{displaymath}
f(r_{\text h}) \equiv \left(r_{\text h} - g_0\beta^0 - \frac{g_0}{(\lambda_\infty^3)^2}\beta_3\right)
\sqrt{(r_{\text h} + 2g_0\beta^0)\left(r_{\text h} + \frac{2g_0}{(\lambda_\infty^3)^2}\beta_3\right)}\,.
\end{displaymath}
In this limit, the spacetime becomes $\mathrm{AdS}_2\times\Sigma$, with metric
\begin{equation}
\mathrm{d}s^2 = -\frac{r^2}{R_1^2}\mathrm{d}t^2 + \frac{R_1^2}{r^2}\mathrm{d}r^2
+ R_2^2\mathrm{d}\Omega_{\kappa}^2\,.
\end{equation}
The Bekenstein-Hawking entropy is given by
\begin{equation}
S_{\text{BH}} = \frac{A_{\text h}}4 = \frac{R_2^2 \,\text{vol}(\Sigma)}4\,.
\end{equation}
This expression can be written in terms of the charges $p^0,q_i$ and the
gaugings $g_0,g^i$ only. To this aim, the eqns.~(\ref{eq:solution})
need to be inverted, in order to use the charges $p^0,q_1,q_2$ as
parameters. This result sustains the presence of the \textit{attractor
mechanism} also in the case under consideration, which is a nontrivial statement, due to
the non-homogeneity of the model we have been discussing.\\
Finally, the product of the areas of all the horizons $r=r_I$, $I=1,\ldots,4$ (i.e., all the roots of
$e^{2U}$) assumes the remarkably simple form
\begin{equation} \label{area-prod}
\prod_{I=1}^4 A(r_I) = -\frac{36}{\Lambda^2}\frac{\text{vol}(\Sigma)^4g^2}{g^3}p^0q_1\tilde q_2^2\,,
\end{equation}
where we defined
\begin{equation}
\tilde q_2 \equiv q_2 - \frac A3\left(\frac{g^3}{g^2}\right)^2 q_1\,.
\label{eq:q_modf}
\end{equation}
Note that \eqref{area-prod} depends only on the charges and the gauge parameters.
Similar formulas have been proven to be true in a number of examples (see
for instance \cite{Raju-Quantum-SKG,Toldo:2012ec,Cvetic:2010mn,Galli:2011fq,Castro:2012av,Klemm:2012vm,Gnecchi:2013mja}),
a fact that calls for an underlying microscopic interpretation.

\section{\label{Conclusion}Conclusions}

In this paper, we considered a non-homogeneous deformation of the stu model
of $N=2$, $D=4$ supergravity, and computed the symplectic embedding of the
electric-magnetic duality algebra. We then focused on a particular FI
gauging of this model, that leads to a scalar potential with two AdS
critical points, a supersymmetric one, and another that breaks supersymmetry
and that exists only when the deformation parameter lies within a specific
range.

Exploiting the construction of this non-homogeneous deformation in string
theory (mentioned at the beginning of section 3), it would be interesting to
investigate the origin of the FI gauging in this context, also in relation
to the $A=0$ limit \cite{Cvetic:1999xp}.

Furthermore, we wrote down the attractor equations for this model, and
constructed an explicit BPS black hole solution that interpolates between
this attractor geometry and the supersymmetric AdS vacuum at infinity.
Various physical properties of this solution were also discussed.

A natural question is whether there exist also black holes in this theory
that asymptotically yield the non-BPS vacuum. Since the first-order flow
equations \eqref{eq:BPS} that we used here are related to the existence of a
Killing spinor \cite{Dall'Agata:2010gj}, they cannot be used to obtain such
non-BPS solutions. A possible way out, that in principle even allows to
construct nonextremal black holes, would be to use the framework of the
Hamilton-Jacobi formalism, leading to first-order equations similar in
spirit to those in \eqref{eq:BPS}.

It would also be interesting to investigate solutions of $\text{AdS}_4$ BPS (and
non-BPS) extremal black holes in $N=2$, $D=4$ FI-gauged supergravity coupled
to hypermultiplets whose quaternionic scalars span non-symmetric (or
non-homogeneous) manifolds, along the lines
of \cite{Halmagyi:2013sla,Erbin:2014hsa,Chimento:2015rra}. In presence of vector
multiplets, a particularly interesting (self-mirror) case consists in the
nh-stu model coupled to four hypermultiplets, whose scalar manifold is the
non-homogeneous c-map image \cite{Cecotti:1988qn} of the non-homogeneous
special K\"{a}hler manifold of the nh-stu model itself.

Finally, non-Abelian gaugings of the vector multiplets' sector (giving rise
to the so-called Einstein-Yang-Mills $N=2$, $D=4$ supergravity theories) are
very little known, especially in relation to the existence and properties of
regular black hole solutions, of the related attractor mechanism, and of
supersymmetry-preserving features. It would be very interesting to study
such issues, e.g. along the lines
of \cite{Huebscher:2007hj,Hubscher:2008yz,Bueno:2014mea,Meessen:2015nla}.

We hope to come back to these points in future publications.

\section*{Acknowledgments}

We would like to thank S.~Cacciatori, S.~Chimento, N.~Halmagyi and M.~Rabbiosi for useful discussions.
This work was partly supported by INFN.

\end{document}